\documentclass{aastex62}

\usepackage[utf8]{inputenc}

\submitjournal{ApJ}

\shorttitle{}
\shortauthors{Yeung et al.}

\begin{document}

\title{The Energy-dependent $\gamma$-ray Morphology of the Crab Nebula Observed with the \emph{Fermi} Large Area Telescope}

\correspondingauthor{Paul K. H. Yeung}
\email{kin.hang.yeung@desy.de}

\author[0000-0003-3476-022X]{Paul K. H. Yeung}
\affil{Institute for Experimental Physics, Department of Physics, University of Hamburg, Luruper Chaussee 149, D-22761 Hamburg, Germany}

\author[0000-0003-1945-0119]{Dieter Horns}
\affil{Institute for Experimental Physics, Department of Physics, University of Hamburg, Luruper Chaussee 149, D-22761 Hamburg, Germany}



\begin{abstract}

The Crab Nebula is a bright emitter of non-thermal radiation across the entire
accessible range of wavelengths. The spatial and spectral structures of the synchrotron nebula are well-resolved from radio to hard X-ray emission. The un-pulsed emission at GeV to TeV energies is mostly produced via inverse-Compton scattering of energetic electrons with the synchrotron-emitted photons. The spatial structure observed at these energies provides insights into the distribution of electrons and indirectly constrains the so-far unknown structure of the magnetic field in the nebula. 
Analyzing
	the LAT data accumulated over $\sim$9.1 years with a properly refined
	model for the Crab pulsar's spectrum, we determined the 68\% containment radius ($R_{68}$) of the Crab Nebula to be $({0.0330\pm0.0025_{stat}}{^{+0.0012}_{-0.0075}}_{sys})^\circ$ (${1.98'\pm0.15'_{stat}}{^{+0.07'}_{-0.45'}}_{sys}$) in the 5--500 GeV band. The estimated systematic
	uncertainty is based on
    two factors: (1) different analysis methods, morphological models and event types, and (2) the point-spread-function evaluated with observations of  Mkn 421. When comparing
	the \emph{Fermi} LAT and H.E.S.S. results on the spatial extension, we find evidence for
	an energy-dependent shrinking of the Crab Nebula's $\gamma$-ray
	extension ($R_{68}\propto  E_\mathrm{IC}^{-\alpha}$ where $\alpha=0.155\pm0.035_{stat}{-0.037}_{sys}$).

\end{abstract}



\section{Introduction} \label{sec:intro}
Isolated neutron stars are efficient particle accelerators, leading to the formation of pulsar wind nebula (PWN) systems. The extended cloud of non-thermal plasma is radiating in a broad energy range, from radio to X-ray and even extending towards the highest gamma-ray energies 
\citep{aharonian_et_al._crab_2004,buhler_surprising_2014,dubner_morphological_2017}.

The Crab Nebula is a PWN powered by a $\sim$1 kyr old pulsar \citep{hester_crab_2008}. 
It is a part of the core-collapse supernova remnant located in the constellation of Taurus and at a distance of 2 kpc \citep{trimble_motions_1968}.
The exceptionally broad energy range  observed from the Crab Nebula enables us to study the processes of particle acceleration occurring at the termination shock \citep[e.g.,][]{spitkovsky_time_2004,fraschetti_particle_2017}.

The discovery of its intense $\gamma$-ray emission dates back to the observations at MeV--GeV energies with the second NASA Small Astronomy Satellite \citep[SAS-2;][]{kniffen_gamma_1974} 
and at TeV energies with the Whipple Observatory 10 m reflector \citep[][]{weekes_observation_1989}. The observed $\gamma$-ray spectrum of the Crab Nebula from 1 GeV to 80 TeV has been compared to various model calculations which use widely different approaches \citep{de_jager_expected_1992,atoyan_mechanisms_1996,hillas_spectrum_1998,volpi_non-thermal_2008,meyer_crab_2010,martin_time-dependent_2012}.
All these models are however based upon the assumption that the gamma-ray emission in this energy range is predominantly produced via inverse-Compton scattering of relativistic electrons with synchrotron-radiated photons as initially suggested by \citet{rees_implications_1971} and \citet{gunn_motion_1971}. The spatial and spectral properties of the synchrotron nebula from optical to $\gamma$-rays are accurately described by a spherically symmetric magnetohydrodynamic model of the outflow forming the Crab Nebula \citep{kennel_magnetohydrodynamic_1984}. The thermal dust emission and the cosmic microwave background contribute to additional seed-photon field.

Investigation into the spatial structure of the Crab Nebula in $\gamma$-ray is certainly required as it will provide additional insights into the concrete mechanisms of its $\gamma$-ray emission. Among different theoretical models, the predicted characteristic size of the $\gamma$-ray nebula varies only a little from 60" \citep[][]{atoyan_mechanisms_1996} to 80" \citep[][]{de_jager_expected_1992}, even though the surface brightness shape is quite different.

In a previous study by \citet{fermi-lat_collaboration_search_2018} with the \emph{Fermi} Large Area Telescope (LAT), an extended morphology seemingly fits the $>$10 GeV $\gamma$-ray emission from the Crab Nebula better than the point model does, even when taking the systematic uncertainties related to the point spread function (PSF) into account. 
On the other hand, the High Energy Stereoscopic System (H.E.S.S.) revealed that the Crab Nebula is extended at TeV $\gamma$-ray energies with a root-mean-square width of 52" \citep{holler_advanced_2017}. 

 The  energy-losses of the electrons diffusing outwards in the nebula lead to the observed energy-dependent size of the synchrotron nebula. Similarly, the $\gamma$-ray extension of the inverse Compton-nebula may  decrease with increasing energy.

In this work, we study the $\gamma$-ray morphology of the Crab Nebula and its energy dependence in detail, with the  $>$5 GeV LAT data accumulated over $\sim$9.1 years and a properly refined model for the Crab pulsar's spectrum. The systematic uncertainties associated with the PSF are evaluated as well. We will compare the physics interpreted respectively from the $\gamma$-ray spectrum, the radio extension and the energy-dependent $\gamma$-ray extension.

\section{Observation \& Data Reduction} \label{sec:data}

We perform a series of unbinned maximum-likelihood analyses for a region of
interest (ROI) of $15^\circ$ radius centered at RA=$05^{h}34^{m}31.94^{s}$,
Dec=$+22^\circ00'52.2"$ (J2000), which is approximately the center of the Crab
Nebula \citep{lobanov_vlbi_2011}. 
We  use the data of $>$5 GeV photon energies,
registered with the LAT between 2008 August 4 and 2017 Sep 25. The data are reduced
and analyzed with the aid of the \emph{Fermi} Science Tools v11r5p3 package.
Considering that the Crab Nebula is quite close to the Galactic plane (with a
Galactic latitude of $-5.7844^\circ$), we adopt the events classified as
Pass8 ``Clean" class for the analysis so as to better suppress the background.
The corresponding instrument response function (IRF) ``P8R2$_-$CLEAN$_-$V6" is
used throughout the investigation. We further filter the data by accepting
only the good time intervals where the ROI was observed at a zenith angle less
than 90$^\circ$ so as to reduce the contamination from the albedo of Earth.

In order to subtract the background contribution, we  include the Galactic diffuse
background (gll$_-$iem$_-$v06.fits), the isotropic background
(iso$_-$P8R2$_-$CLEAN$_-$V6$_-$v06.txt) as well as all other  point sources
cataloged in the most updated Fermi/LAT catalog (FL8Y~\footnote{\label{FL8Y_ref} Fermi-LAT 8-year Source List: \url{https://fermi.gsfc.nasa.gov/ssc/data/access/lat/fl8y}})
within 25$^\circ$ from the ROI center in the source model.  We  set free the
spectral parameters of the  sources within 5$^\circ$ from the ROI center in the
analysis. For the   sources beyond 5$^\circ$ from the ROI center, their
spectral parameters are fixed to the catalog values. 

The two point sources located within the nebula are cataloged 
as  FL8Y J0534.5+2200 and FL8Y J0534.5+2201i, which respectively model the
Crab pulsar and  the Crab Nebula. We leave the point-source morphology of the
pulsar component unchanged throughout our work.  For  the PWN component, we
choose a point-source model as well as disk models of different radii, in order to
determine the most likely morphology in each energy range we chose.

\begin{figure}
	\plotone{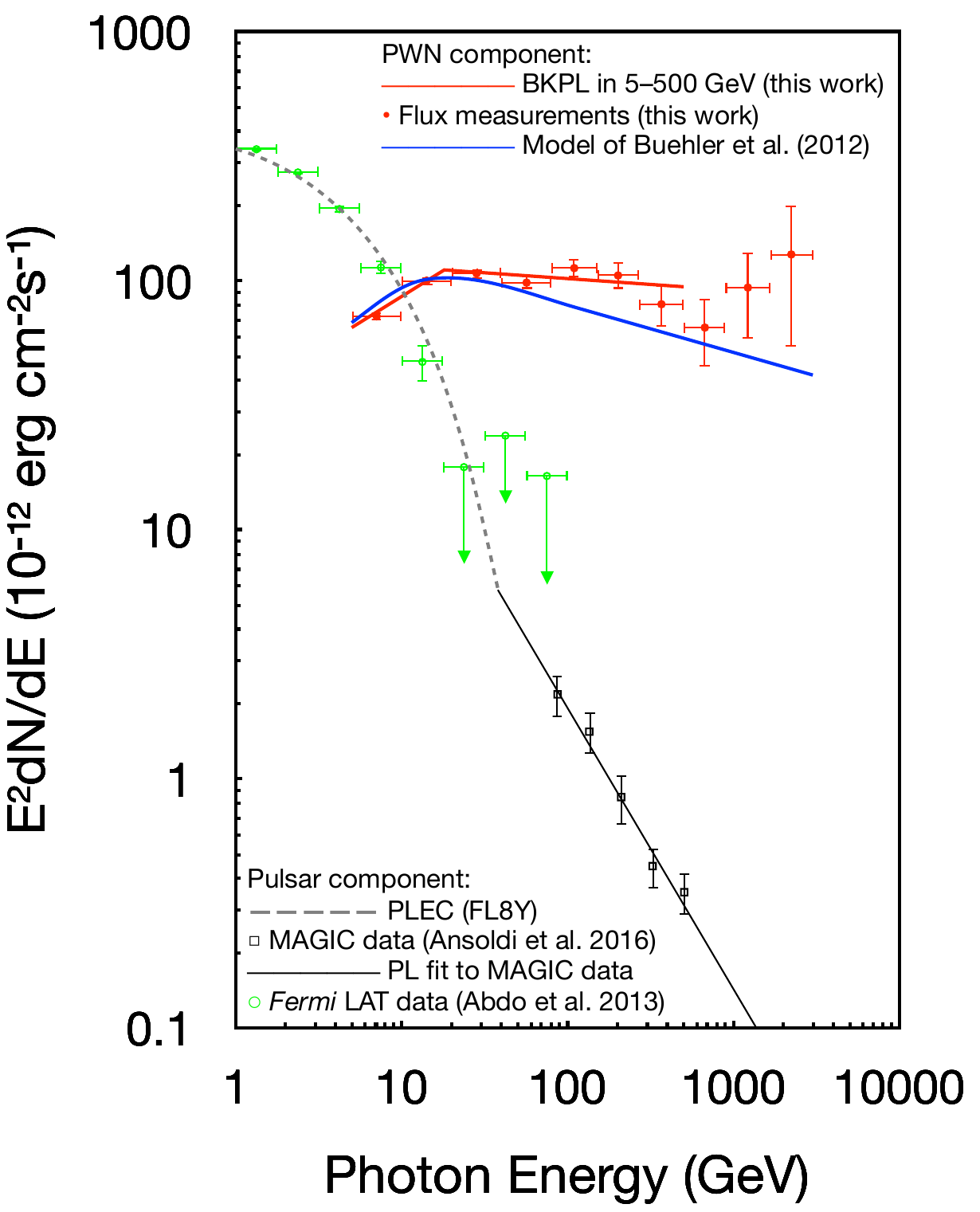}
	\caption{The GeV--TeV spectral energy distribution of the Crab pulsar (in gray, black and green) and the Crab  PWN (in red and blue). The gray dashed curve represents the PLEC pulsar model in FL8Y, which is kept for  energies below 38 GeV. { The phase-averaged pulsar spectrum as reported in the LAT Second Pulsar Catalog \citep{abdo_second_2013} is shown as green open circles for comparison.} The black solid line represents the  PL model fit to the  69--628 GeV pulsar spectrum measured with  MAGIC (open squares), where the data is taken from  \citet{ansoldi_teraelectronvolt_2016}. It replaces the PLEC for   energies above 38 GeV. The red dots are the \emph{Fermi} LAT fluxes of the PWN determined in our analyses. The red line represents the maximum-likelihood  broken-power-law (BKPL) model we determined for the 5--500 GeV  PWN spectrum (see \S\ref{EED} for more detail). The blue solid curve represents the off-pulse model of the PWN reported by \citet{buehler_gamma-ray_2012}.  \label{PSR_LAT-MAGIC}} 
\end{figure}

We fix the spectral parameters of FL8Y J0534.5+2200 (the pulsar
component) at certain values so as to avoid degeneracies in the fitting procedure.
Since it is the most contaminating `background' source in our work and
its spectral fitting of a power law with a sub-exponential cutoff (PLEC) in the FL8Y catalog is dominated by the $<$1 GeV data,  we
refine its spectral model at larger energies based on the phase-folded spectrum of the Crab
pulsar in 69--628 GeV measured with 
MAGIC \citep{ansoldi_teraelectronvolt_2016}. We thereby determine a  power-law (PL)
spectrum  (see Figure~\ref{PSR_LAT-MAGIC}): 
\begin{center}
	$\frac{dN}{dE}=2.19\times10^{-10}(\frac{E}{GeV})^{-3.13}$ photons cm$^{-2}$ s$^{-1}$ MeV$^{-1}$ \ \ \ .
\end{center}
This PL model intersects with the catalog PLEC model at $\sim$38 GeV, below which the PL seriously under-predicts the Crab pulsar's flux. {  Below 38 GeV, the more recent PLEC model (FL8Y) is in a good agreement with the binned spectrum reported in the LAT Second Pulsar Catalog \citep{abdo_second_2013} - see also Fig.~\ref{PSR_LAT-MAGIC}.} Therefore, we keep the FL8Y  spectrum of the Crab pulsar for energies below 38 GeV, while we replace the PLEC with the PL for other energies. 

As presented in Figure~\ref{PSR_LAT-MAGIC} and  \S\ref{EED}, such a hybrid model for FL8Y J0534.5+2200 (the pulsar component) yields a $>$5 GeV spectrum of FL8Y J0534.5+2201i (the PWN component) which essentially matches the off-pulse spectrum reported by \citet{buehler_gamma-ray_2012}. { In particular, above 20 GeV (and 40 GeV), the predicted flux of the Crab pulsar only accounts for $\le21\%$ (and $<5\%(E/40~GeV)^{-1}$) of the Crab system's total flux. Clearly, the spectral model assigned to the Crab pulsar is not expected to introduce any obvious bias in our analyses for such high energies, due to its minor contribution of flux.}

\section{Data Analysis \& Results}

\subsection{The centroid of the nebula emission} \label{centroid}

\begin{figure}
	\plotone{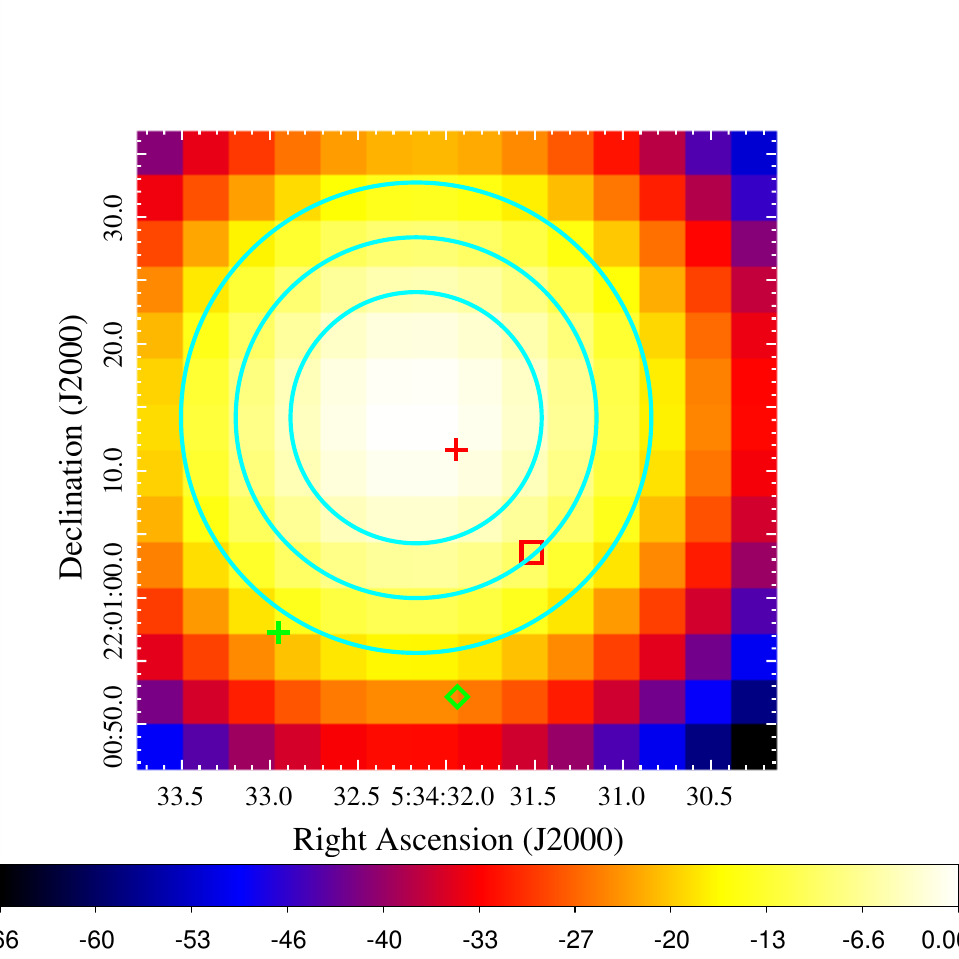}
	\caption{$\delta$TS map of the field around the Crab system in 5--500 GeV,
	where all FL8Y catalog sources except  FL8Y J0534.5+2201i (the Crab
	PWN)  are subtracted. The color scale represents the TS value subtracting the maximum. The pixel size
	($0.001^\circ\times0.001^\circ$) oversamples the PSF and the map covers a  field of view of $0.014^\circ\times0.014^\circ$. The green and red crosses represent
	the catalog positions of  FL8Y J0534.5+2200 (the Crab pulsar) and  FL8Y
	J0534.5+2201i respectively. The 68\%, 95\% and 99.7\% error circles of
the $\gamma$-ray centroid for four degrees of freedom, where the TS value is
lower than the maximum by 4.7, 9.5 and 16.0
respectively~$^{\ref{error_circle}}$, are plotted in cyan (from innermost to
outermost). The red square indicates the radio centroid of the Crab Nebula
which is determined from a VLA (5.5~GHz) image published in
\citet{bietenholz_crab_2004} (see \S\ref{DiscuRadio} and Figure~\ref{VLA} for more
detail). The green diamond indicates the radio position of the Crab pulsar
taken from \citep{lobanov_vlbi_2011}. \label{fullband_tsmap}} 
\end{figure}

The 5--500 GeV test-statistic (TS) map is shown in Figure~\ref{fullband_tsmap},
where all FL8Y catalog sources except  FL8Y J0534.5+2201i (the PWN component of
the Crab system) are subtracted. The pixel size of the map is chosen that   the PSF is oversampled
($0.001^\circ\times0.001^\circ$). Therefore, the map covers a small field of view
($0.014^\circ\times0.014^\circ$). The TS map demonstrates that the catalog position of
FL8Y J0534.5+2201i (marked as a red cross in Fig.~\ref{fullband_tsmap}) is comfortably located within the 68\% error circle of the centroid
for four degrees of freedom (d.o.f.), where the TS value is lower than the
maximum by 4.7~\footnote{\label{error_circle}A $\chi^2$ Distribution is
assumed. There are four d.o.f. because of the four variables: R.A., decl., flux
normalization, and photon index.}. This centroid is within $2\sigma$ consistent with the 
centroid position of the radio nebula (marked as red box in Fig.~\ref{fullband_tsmap}) but
offset from the radio
position of the Crab pulsar at a $>$$3\sigma$ level.

\begin{figure}
	\plotone{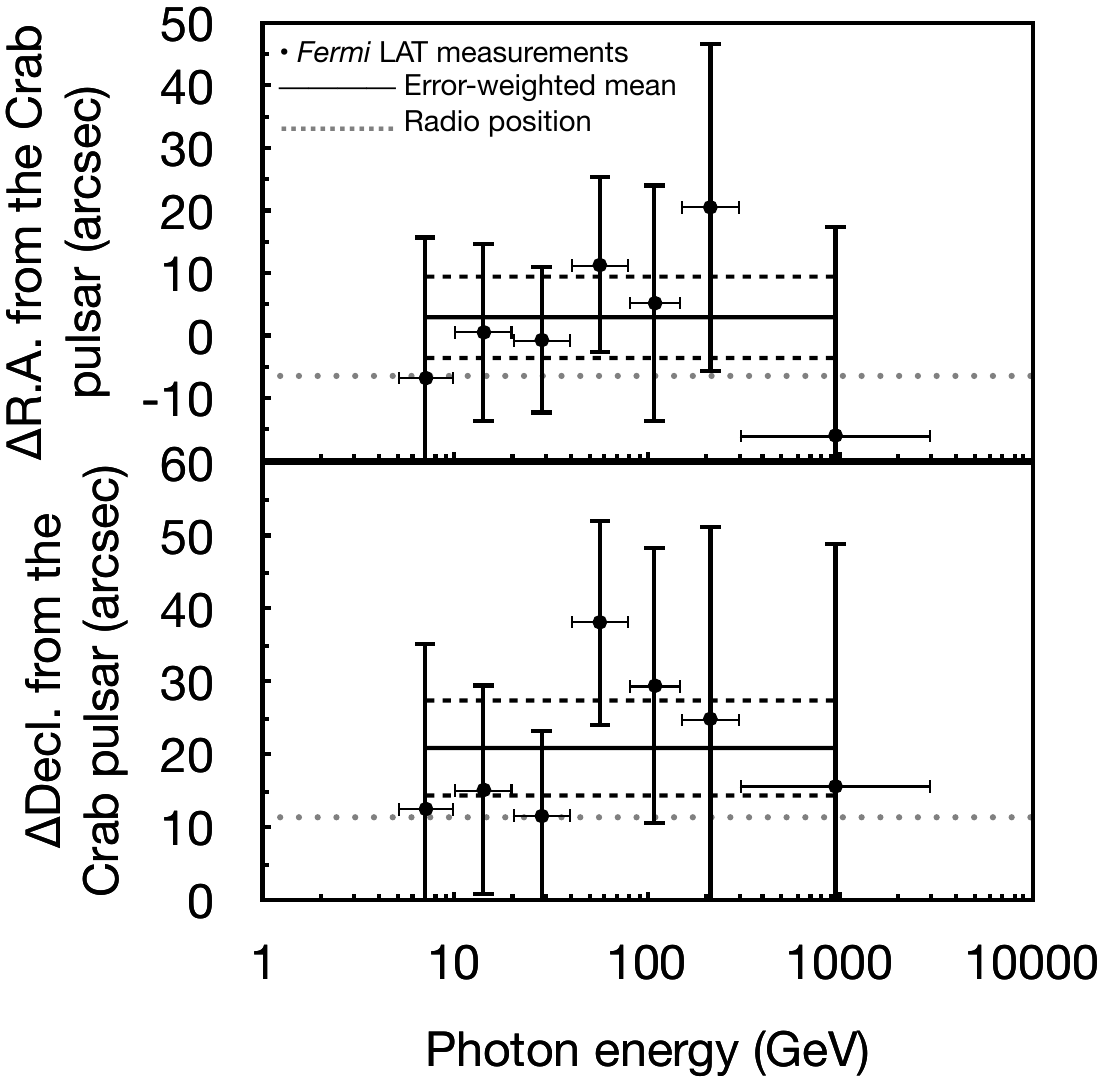}
	\caption{Difference of the Crab Nebula's  centroid position from the
radio position of the Crab pulsar taken from \citep{lobanov_vlbi_2011} in
different energy-segments, along the axes of right ascension and declination
respectively. On each panel, the black solid line indicates the best-fit
constant-value function (i.e., the error-weighted mean), and sandwiched between the black dashed lines is its
$1\sigma$ error range. The gray dotted lines indicate the position of the Crab
Nebula's radio centroid (determined from Figure~\ref{VLA}) relative to the
radio position of the Crab pulsar. \label{centroids}} 
\end{figure}

We divide the entire  5--3000 GeV band into several energy intervals: 5--10 GeV,
10--20 GeV, 20--40 GeV, 40--80 GeV, 80--150 GeV, 150--300 GeV, and 0.3--3
TeV~\footnote{The spectral coverages of the Galactic diffuse background
(gll$_-$iem$_-$v06.fits) and the isotropic background
(iso$_-$P8R2$_-$CLEAN$_-$V6$_-$v06.txt) are up to $\sim$0.5 TeV and $\sim$0.9
TeV respectively. Yet, their contamination becomes  negligible above 0.3
TeV. Therefore, we remove them from the source model for  the
0.3--3 TeV analyses.}. For each spectral segment, we repeated creating the TS
map with  the same pixel size and  field of view. The separations of Crab
Nebula's centroids from the radio position of the Crab pulsar are plotted in
Figure~\ref{centroids}. The centroids in all the seven segments are
consistently offset from the pulsar by ${\Delta}R.A.=(3.05\pm6.51)"$
($\chi^2\sim1.41$ for 6 d.o.f.) and ${\Delta}Decl.=(20.86\pm6.51)"$
($\chi^2\sim2.60$ for 6 d.o.f.).

Since \emph{no} discrepancy among the centroids in different energy bands and
the  FL8Y position can be robustly claimed, we consistently leave the
position of  FL8Y J0534.5+2201i unchanged in subsequent analyses (i.e., there are no
additional degrees of freedom from the centroid position).

\subsection{Variability of the flux}

\begin{figure}
	\plotone{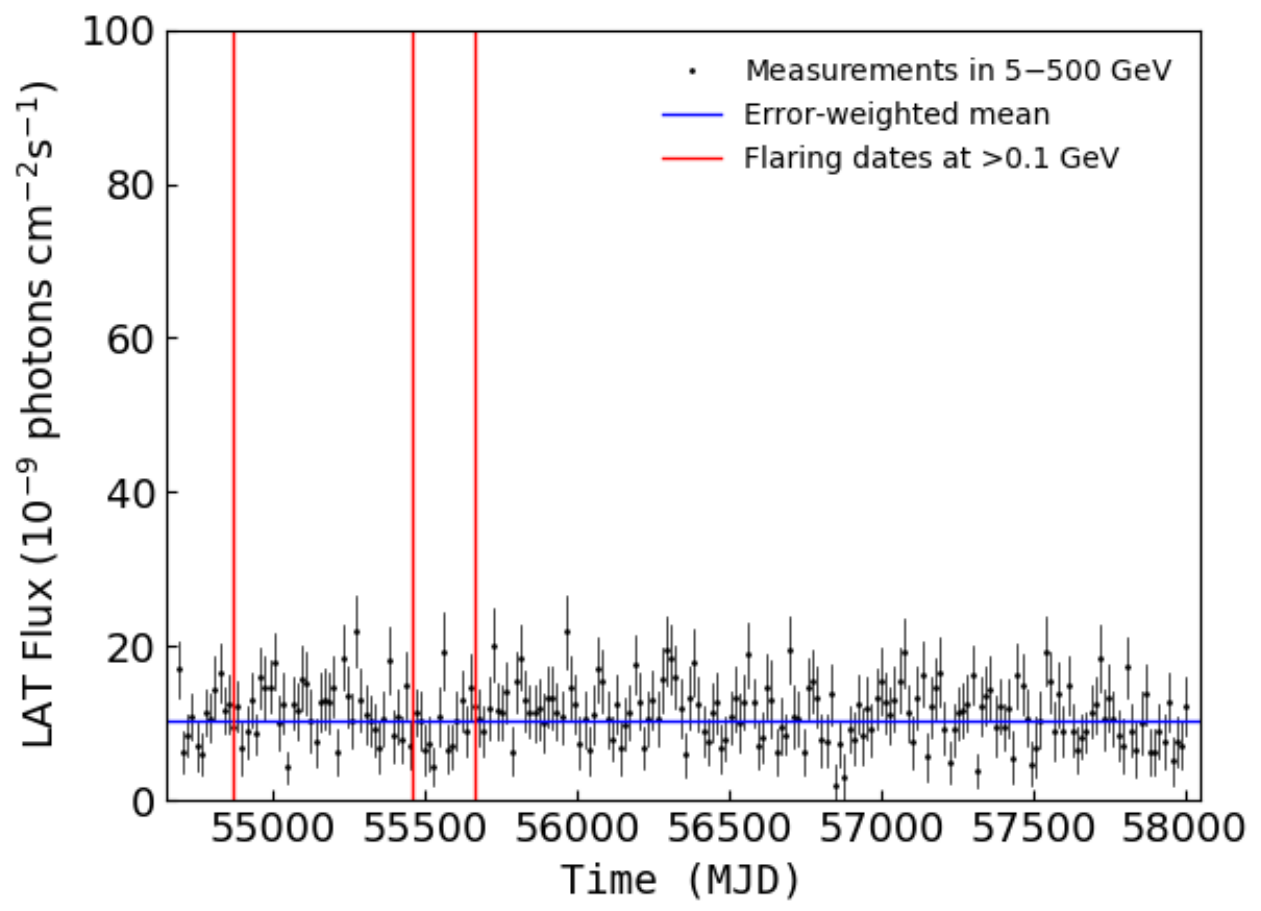}
	\caption{5--500 GeV light-curve of FL8Y J0534.5+2201i, with a bin size of 15 days.  The red vertical lines indicate the dates of the flares detected at $>$0.1 GeV by \citet{buehler_gamma-ray_2012}. The blue solid line indicates the best-fit constant flux (i.e., the error-weighted mean).\label{Crab_lc}}
\end{figure}

We divide the first $\sim$9.1 years of \emph{Fermi} LAT observation into a
number of 15-day segments, and perform an unbinned maximum-likelihood
analysis for 5--500 GeV data in each individual temporal segment. Considering that the
isotropic background $\gamma$-ray emission \emph{cannot}  noticeably change
within a short timescale of 10 years, we  fix it at the
$\sim$9.1-year average obtained from the full-timespan analysis, so that the
statistical fluctuations  from the isotropic background model are
avoided. The light-curve of  FL8Y J0534.5+2201i is shown in
Figure~\ref{Crab_lc}.

A constant flux satisfactorily fits the
entire temporal distribution  with $\chi^2\sim256$ for 221 d.o.f. ($p(>\chi^2)=0.05$). In order to check for significant deviations where subsequent flux points are either above or below the average,  we perform an additional Wald-Wolfowitz run test (where we define two kinds of runs: runs of bins above the
average and runs of those below it), the observed number of runs deviates from
the expected number by only $\sim0.6\sigma$. Furthermore, the 5--500 GeV flux
of the nebula shows no correlation with the flares which enhanced its $>$0.1 GeV flux by a factor of $>$5 \citep[cf. Figure~3 of][]{buehler_gamma-ray_2012}. This is as expected because the $\gamma$-ray spectra
during the flaring states have their cutoff energies well below 1 GeV
\citep[cf. Figures~6~\&~7 of][]{buehler_gamma-ray_2012}. 

We hereby confirm that the $\gamma$-ray flares of the Crab system at lower
energies do not perturb the results above 5 GeV and the 5--500 GeV flux
is essentially steady. It is, therefore, appropriate to accept all the good
time intervals between 2008 August 4 and 2017 Sep 25 in subsequent analyses
(i.e., no further screening of data is required).

\subsection{Extension \& its energy-dependence} \label{EED}

\begin{figure}
	\plotone{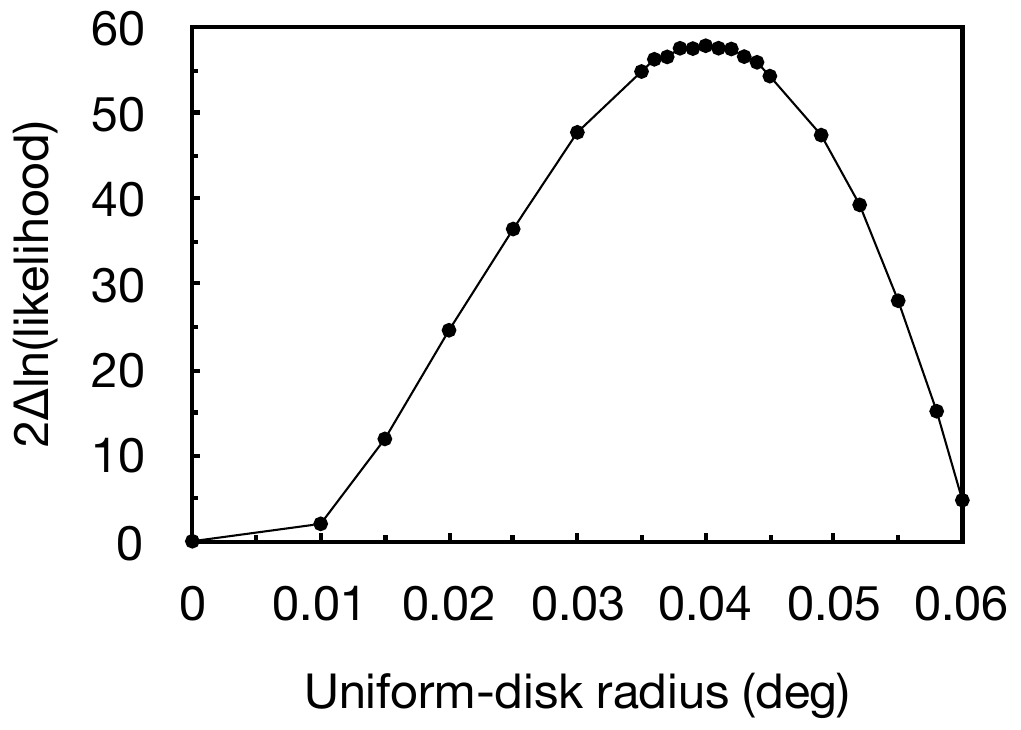}
	\caption{The $2\Delta$ln(likelihood) in 5--500 GeV, when uniform disks of different radii replace the point-source model to be the morphology of FL8Y J0534.5+2201i. \label{Ext}}
\end{figure}

In order to examine whether the   $\gamma$-ray emission from the
PWN is spatially extended, we  perform a likelihood-ratio test to quantify the
significance of extension in the 5--500 GeV band. After refinement of the Crab pulsar's spectrum, we found that a broken-power-law (BKPL)  spectral model is  preferred over a PL by $\sim5.4\sigma$ for the PWN ($\Delta{TS}\sim32.7$ for 2 d.o.f.). The spectrum of the PWN softens from $\Gamma_{1}=1.585 \pm 0.067$ to $\Gamma_{2}=2.047 \pm 0.036$ at $E_\mathrm{b}=18.05 \pm 1.60$ GeV, consistently with the spectral model determined for the off-pulse phase by \citet{buehler_gamma-ray_2012} (see Figure~\ref{PSR_LAT-MAGIC}). Therefore, we assign it a BKPL spectral model. We attempt uniform-disk morphologies
of different sizes as well as a point-source model on it. The
$2\Delta\ln(likelihood)$ of different sizes relative to the point-source model
are plotted in Figure~\ref{Ext}.

The most likely disk radius is determined to be
$(0.040\pm0.003)^\circ$  and this morphology is preferred
over a point-source model by $\sim7.6\sigma$. The corresponding 68\% containment radius ($R_{68}$; the uniform-disk radius multiplied by $\sqrt{0.68}$) of $(0.0330\pm0.0025)^\circ$  resp. $(1.98\pm0.15)'$ is consistent with that determined in \citet{fermi-lat_collaboration_search_2018} with 1~GeV--1~TeV data and a Gaussian morphology ($0.030^\circ\pm0.003^\circ_{stat}\pm0.007^\circ_{sys}$), within the tolerance of statistical uncertainties. Motivated by the uncertainties in the Crab pulsar's spectrum, we repeated this analysis with altering the flux normalization of the Crab pulsar by $\pm$20\%. It turns out that the maximum-likelihood radius remains unchanged even though $\Gamma_{1}$ is altered by $^{+0.28}_{-0.43}$ (i.e., the spectral model of the Crab pulsar has no noticeable contribution to the systematic uncertainty).

We further verified the robustness of the 5--500 GeV result by performing binned maximum-likelihood analyses with the aid of the ``fermipy"  package~\footnote{\label{fermipy_ref}Fermipy's documentation: \url{https://fermipy.readthedocs.io/en/latest}} \citep{Wood_Fermipy_2017}. We adopted a bin size of $0.01^\circ$ which is sufficiently small to well sample the PSF as well as the $\gamma$-ray nebula. In addition to the analysis with ``FRONT+BACK" data, we also performed a joint analysis with ``PSF2" and ``PSF3" data, and an analysis with only ``PSF3" data (respectively sacrificing the photon statistics by a factor of $\sim$1/2 and $\sim$1/4 for better spatial resolution). For each data set we worked on, we examined both uniform-disk and Gaussian morphologies. 

As can be seen in Table~\ref{fermipy}, regardless of the event type and morphological model, the values of $R_{68}$ are all consistent with $0.0330^\circ\pm0.0025^\circ$ (the result of the unbinned maximum-likelihood analysis) within the tolerance of statistical uncertainties. For each event type we attempted, the two morphological models have roughly the same goodness of fit ($\Delta TS_{ext}\leq 1.4$), and their difference in $R_{68}$ is negligible ($\leq0.6\sigma$). Also, screening out the data partitions of poorer resolution did not lead to a noticeable drop in $TS_{ext}$. We hereby  compute a systematic uncertainty of $R_{68}$ of $\pm0.0012^\circ$ which stems from  the analysis method, morphological model and event selection.

\begin{table}
	\centering
	\caption{Morphological studies for FL8Y J0534.5+2201i in 5--500 GeV with different analysis methods, morphological models and event types. }
	\label{fermipy}
\begin{tabular}{l|c|ccc}\hline\hline
Event type & Morphological model & Radius (deg) & $R_{68}$ (deg)~$^\mathrm{a}$ & $TS_{ext}~^\mathrm{b}$ \\ \hline
\multicolumn{5}{c}{Unbinned maximum-likelihood analysis}     \\ \hline
FRONT+BACK & Disk                & $0.040\pm0.003$       & $0.0330\pm0.0025$            & 57.81  \\ \hline
\multicolumn{5}{c}{Binned maximum-likelihood analysis in ``fermipy",  bin size = $0.01^\circ$}     \\ \hline
FRONT+BACK & Disk                & $0.0385^{+0.0033}_{-0.0036}$       & $0.0317^{+0.0028}_{-0.0030}$            & 45.50  \\
           & Gaussian            & --            & $0.0307^{+0.0029}_{-0.0030}$            & 45.59  \\ \hline
PSF2+PSF3  & Disk                & $0.0400^{+0.0028}_{-0.0032}$       & $0.0330^{+0.0023}_{-0.0027}$            & 62.34  \\
           & Gaussian            & --            & $0.0308^{+0.0025}_{-0.0026}$            & 60.93  \\ \hline
PSF3  & Disk                & $0.0405^{+0.0033}_{-0.0036}$       & $0.0334^{+0.0027}_{-0.0030}$            & 51.48  \\
           & Gaussian            & --            & $0.0311^{+0.0028}_{-0.0029}$            & 50.54  \\ \hline
\end{tabular}

\raggedright
$^\mathrm{a}$ The 68\% containment radius. For a disk model, it is the radius  multiplied by $\sqrt{0.68}$.\\
$^\mathrm{b}$ The $2\Delta\ln(likelihood)$ between the best-fit morphology and the point-source model.\\

\end{table}

In order to investigate whether the $\gamma$-ray morphology  changes with 
photon energy, we divide the entire  5--3000 GeV band in the same way
as  in \S\ref{centroid}. We repeat  the likelihood ratio test
for each spectral segment, with a PL assigned to the PWN spectrum.
The results are tabulated in Table~\ref{dependence}. We sum up the
differences between $2\ln(L_{ext,max}/L_{pt})$ and $2\ln(L_{ext,0.04}/L_{pt})$
over the all seven segments, and hence we get a  TS value of the
energy-dependence of 18.2 for 7 d.o.f.. In other words, based on our
\emph{Fermi} LAT results only, an energy-dependent morphology with the nebula size shrinking with increasing energy is preferred over
a constant size by  $\sim2.5\sigma$. In addition, the PWN's flux in each segment is well consistent with the off-pulse spectrum reported by \citet{buehler_gamma-ray_2012}. This confirms that the systematic uncertainties associated with the Crab pulsar's spectral model are not a serious issue.

\begin{table}
	\centering
	\caption{Morphological studies for the Crab PWN in different energy-segments, with \emph{Fermi} LAT and H.E.S.S.. }
	\label{dependence}
	\begin{tabular}{ccccc} \hline\hline
		Energy range (GeV) & Disk radius (deg)                  & $R_{68}$ (deg)~$^\mathrm{a}$ & $2\ln(L_{ext,max}/L_{pt})$~$^\mathrm{b}$ & $2\ln(L_{ext,0.04}/L_{pt})$~$^\mathrm{c}$ \\ \hline
\multicolumn{5}{c}{\emph{Fermi} LAT result}     \\ \hline
		5--500               & $0.040\pm0.003$ & $0.0330\pm0.0025$ &         57.81       &      57.81           \\  \hline
		5--10               & $0.073^{+0.010}_{-0.011}$ & $0.0602^{+0.0082}_{-0.0091}$ &         15.75       &      8.40           \\
		10--20              & $0.057^{+0.005}_{-0.006}$ & $0.0470^{+0.0041}_{-0.0049}$ &         33.80       &      26.27           \\
		20--40              & $0.034^{+0.007}_{-0.005}$ & $0.0280^{+0.0058}_{-0.0041}$ &         11.14       &      10.52           \\
		40--80              & $0.038^{+0.005}_{-0.006}$ & $0.0313^{+0.0041}_{-0.0049}$ &         13.78       &      13.46           \\
		80--150             & $0.032^{+0.010}_{-0.008}$ & $0.0264^{+0.0082}_{-0.0066}$ &         5.06       &      4.40           \\
		150--300            & $0.028^{+0.009}_{-0.010}$ & $0.0231^{+0.0074}_{-0.0082}$ &         3.34       &      1.63           \\
		300--3000            & $0.041^{+0.013}_{-0.016}$ & $0.0338^{+0.0107}_{-0.0132}$ &         2.78       &      2.76           \\ \hline
\multicolumn{5}{c}{H.E.S.S. result \citep{holler_advanced_2017}}     \\ \hline
		700--10000            & -- & $0.0219\pm0.0012_{stat}\pm0.0033_{sys}$ &        83       &      --           \\ \hline
	\end{tabular}
	
	\raggedright
	$^\mathrm{a}$ The 68\% containment radius, which is the disk radius  multiplied by $\sqrt{0.68}$.\\
	$^\mathrm{b}$ The $2\Delta\ln(likelihood)$ between the best-fit uniform-disk morphology and the point-source model.\\
	$^\mathrm{c}$ The $2\Delta\ln(likelihood)$ between the uniform-disk morphology of a $0.04^\circ$ radius and the point-source model.\\
\end{table}

\begin{figure}
	\plotone{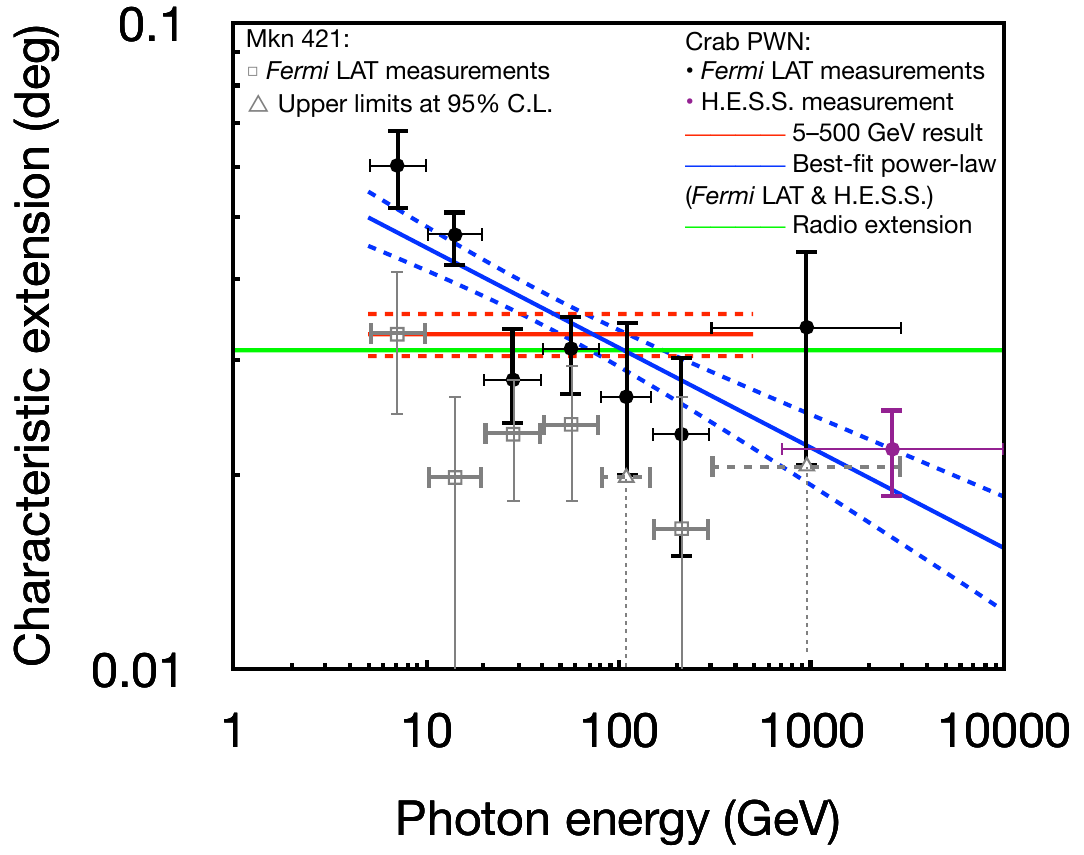}
	\caption{Characteristic  extensions (defined as $R_{68}$; for a uniform-disk model, it is the disk radius multiplied by $\sqrt{0.68}$) of the Crab PWN in different energy-segments, from the \emph{Fermi}
	LAT band to the  H.E.S.S. band (in which the data is taken from
	\citet{holler_advanced_2017}). For the H.E.S.S. bin, we take its combined uncertainty (where  statistical
	and systematic uncertainties are added in quadrature).  The red solid line indicates the characteristic extension of
	the energy-independent morphology (disk) fit to the 5--500 GeV
	emission, and sandwiched between the red dashed lines is its $1\sigma$
	error range. The blue line is the power-law function which best
	describes the relation of characteristic extension to the photon
	energy, and sandwiched between the blue dashed lines is its $1\sigma$ error range.  The green solid line indicates the characteristic radio
	extension, based on a VLA (5.5~GHz) image
	published in \citet{bietenholz_crab_2004} (see \S\ref{DiscuRadio} and
	Figure~\ref{VLA} for more detail about how the measurement is done).
	The `apparent' extensions of Mkn 421 determined through the same
	procedures (for evaluating the PSF) are plotted in gray, and we place upper limits of a 95\%
confidence level for those segments in which a point model is preferred over
any  uniform-disk model. \label{FermiHESS}} 
\end{figure}

We scale the disk radii determined with LAT to $R_{68}$, so that they can be compared with the H.E.S.S. Gaussian extension reported by \citet{holler_advanced_2017}, as plotted in Figure~\ref{FermiHESS}. The $R_{68}$ in 5--500 GeV observed with \emph{Fermi} LAT is larger than that observed at higher energies of 0.7--10 TeV with H.E.S.S. at a $\sim2.6\sigma$ level. A constant-value function  yields a poor fit to the distribution of $R_{68}$ ($\chi^2\sim29.4$ for 7 d.o.f.; i.e., it deviates from a uniform distribution at a $\sim3.8\sigma$ level). When we fit a power-law function instead, the goodness of fit greatly improves ($\chi^2\sim8.7$ for 6 d.o.f.). An F-test yields a statistic of $\sim14.3$ for (1,6) d.o.f., implying a chance probability of $\le$0.9\% for the energy-dependent shrinking.  Thus, the dependence of $R_{68}$ on the photon energy (E) can be formulated as 
\begin{center}
	$R_{68}=(0.0357\pm0.0021)(\frac{E}{44.0~\mathrm{GeV}})^{-0.155\pm0.035}$ deg
\end{center}
without any correlation between the prefactor and index.

{ The extension of the nebula at energies between 20 GeV and 40 GeV appears to deviate from the power-law ($\sim1.8\sigma$). Interestingly, the energy flux of the nebula levels off to an almost flat peak at the same energy (see Figure~\ref{PSR_LAT-MAGIC}).}

\section{Discussion}

\subsection{Evaluation of the PSF}

Since the radius of the most likely uniform-disk morphology of FL8Y
J0534.5+2201i is at least two times smaller than  the 68\% containment radius
of the acceptance weighted PSF for all energy bands we investigate (see
SLAC~\footnote{\label{slac}Fermi LAT Performance:
\url{http://www.slac.stanford.edu/exp/glast/groups/canda/lat_Performance.htm}}),
it is necessary to evaluate the accuracy of the IRF ``P8R2$_-$CLEAN$_-$V6" we
used. We did this by determining the `apparent' $\gamma$-ray extension of Mkn
421, a GeV- and  TeV-bright  blazar at high Galactic
latitude, through the same procedures.

It turns out that the best-fit uniform-disk morphology of  Mkn 421 in 5--500
GeV has a radius of $0^\circ.025^{+0^\circ.003}_{-0^\circ.004}$ with $TS_{ext}$ of 10.2, which are
significantly smaller than those determined for the Crab Nebula. Also, the most
likely extensions of  Mkn 421 in the divided energy-segments are, as overlaid
in Figure~\ref{FermiHESS}, collectively smaller than those of the Crab Nebula.
Assuming that the extension determined for Mkn~421 is purely an instrumental effect, we estimate a systematic uncertainty of $-0.009^\circ$ for the disk radius of  the Crab Nebula in 5--500 GeV (corresponding to a systematic uncertainty of $-0.0074^\circ$ for $R_{68}$). Combining this with the effects of changing the analysis method, morphological model, event type and Crab pulsar's spectrum, we estimate the total systematic uncertainty of $R_{68}$ to be $(^{+0.0012}_{-0.0075})^\circ$.

\subsection{Comparison of the $\gamma$-ray nebula to the radio nebula} \label{DiscuRadio}

\begin{figure}
		\plotone{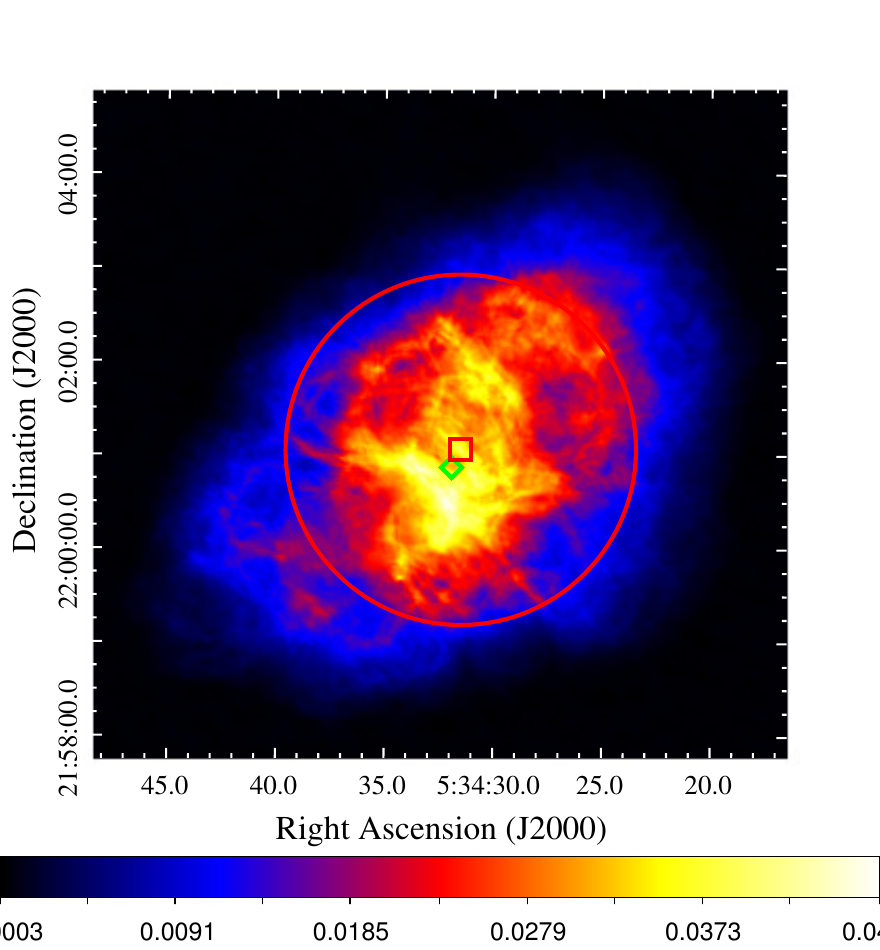}
	\caption{VLA radio (5.5~GHz) image of the Crab Nebula retrieved from \citet{bietenholz_crab_2004}. The red square indicates the intensity-weighted centroid, which is also overlaid in Figure~\ref{fullband_tsmap}. The green diamond indicates the radio position of the Crab pulsar taken from \citet{lobanov_vlbi_2011}. The position of the intensity-weighted centroid relative to the Crab pulsar is indicated in Figure~\ref{centroids}. The red circle centered at the centroid indicates the 68\% containment radius which is overlaid in Figure~\ref{FermiHESS}. \label{VLA}}
\end{figure}

In order to compare the $\gamma$-ray morphology with the radio morphology for
the Crab Nebula, we retrieved a VLA (5.5~GHz) image (Figure~\ref{VLA}) published in
\citet{bietenholz_crab_2004}. First of all, we determined the
intensity-weighted centroid and overlaid it in Figure~\ref{fullband_tsmap}. The
position of this centroid relative to the Crab pulsar is
indicated in Figure~\ref{centroids}. 
Then, we determined the 68\% containment circle centered at this centroid and its radius is overlaid in Figure~\ref{FermiHESS}.

As shown, the radio centroid is on the very edge of the 95\% error circle of the 5--500 GeV
centroid. Both the $\gamma$-ray and radio
centroids of the PWN are northward offset from the Crab pulsar. Neglecting the systematic uncertainties associated with the PSF, the average of 5--20 GeV
extensions is  larger than the radio extension  by $\sim4.7\sigma$, while the $>$20 GeV
extensions do not exceed the  radio extension. Even after reducing the 5--20 GeV extensions based on the assumption that the LAT extensions determined for Mkn~421 are purely instrumental effects, their average still exceeds the radio extension  by $\sim2.9\sigma$.

The comparison of the size of the inverse Compton nebula at 5--20 GeV 
with the size of the synchrotron nebula at 5~GHz provides a measure of the ratio of seed photon field energy density and magnetic field energy density. The synchrotron emission at 5 GHz is mainly produced by electrons with Lorentz factors $\gamma \approx 6\times 10^3 (B/120~\mu\mathrm{G})^{-1/2}$. The same electrons will produce inverse Compton emission at energies below a GeV. Therefore, the ratio of $r_{IC}/r_{Sy}$ is a measure of the 
size of the seed photon field $r_{seed}$ and the size of the magnetized nebula $r_{B}$. In a more detailed modeling approach, it is therefore necessary to include the spatial distribution of additional seed photon fields, including the emission of the dusty plasma in which the synchrotron nebula is embedded. 

\subsection{Comparison of the observed energy-dependence of the $\gamma$-ray extension to theoretical models}

We  found that, as the photon energy ($E_\mathrm{IC}$) increases from 5 GeV to  10
TeV, the spatial extension of the Crab Nebula is shrinking with a power-law index such that $R_{68}\propto  E_\mathrm{IC}^{-0.155\pm0.035}$. Even after we modified the spectral distribution of Crab Nebula's extensions  based on the assumption that the LAT extensions determined for Mkn~421 are purely instrumental effects,  the size of the Crab Nebula still deviates from a uniform distribution at a $\sim2.9\sigma$ level, and it still shrinks with a power-law index of $\sim0.118$ which is a reasonable estimate of the systematic lower bound. Such an observed energy-dependence of the $\gamma$-ray extension is comparable to the energy dependence of the size of the underlying electron distribution which was found to be $r_\mathrm{e}\propto \gamma^{-0.17}$ \citep{meyer_crab_2010} when assuming a homogeneous magnetic field. 
{This approach effectively models the radiative cooling of electrons while expanding into the nebula.}

For  Thomson-type inverse-Compton scattering with $E_\mathrm{IC}\approx \gamma^2\epsilon$ (where $\epsilon$ is the seed photon energy), provided that the spectral number density of the seed photon field is uniform (e.g., like the cosmic microwave background (CMB)), the resulting nebula size should shrink with a  harder power-law $R_{68}\propto \sqrt{r_e} \propto E_\mathrm{IC}^{-0.17/2}$ which is
similar to the  energy-dependence of the synchrotron nebula size. 
{However, at energies larger than a few 100~GeV, inverse Compton scattering with the synchrotron seed photon field starts to be affected by Klein-Nishina effects. In the case of dominating Klein-Nishina effects, the energy dependence of the inverse Compton nebula will proceed with $R_{68}\propto r_e \propto E_{IC}^{-0.17}$. Even though this is apparently a closer match to the observed energy dependence,  Klein-Nishina effects are not expected to dominate in the  low-energy part ($E<500$~GeV) covered with the measurement presented here. 
}

{The stronger energy-dependence}  can be interpreted  through 
a change in the ratio of energy densities $u_*(r)/u_B(r)$ in the seed photon field $u_*$ and in the magnetic field $u_B$. In turn,  this may be an indication of an unknown magnetic field structure in the nebula. 
{Further details on the interpretation of the energy-dependence of the spatial extent of the inverse Compton-nebula require a  careful  modelling of the interplay of the spatial distribution of seed photon and magnetic fields and of the transition between Thomson and Klein-Nishina scattering, which is beyond the scope of this publication.}

\section{Summary}

With the proper refinement of the spectral model of the Crab pulsar, we unbiasedly determined the 68\% containment radius ($R_{68}$) of the inverse-Compton nebula to be $({0.0330\pm0.0025_{stat}}{^{+0.0012}_{-0.0075}}_{sys})^\circ$ (${1.98'\pm0.15'_{stat}}{^{+0.07'}_{-0.45'}}_{sys}$) in the 5--500 GeV band. The particularly large 5--20 GeV extensions compared with the radio size of the synchrotron nebula implies that additional sources of seed photons (e.g., CMB and dust) must be taken into account in theoretical modeling. 
The strong energy-dependence of its extension from 5 GeV to  10 TeV ($R_{68}\propto  E_\mathrm{IC}^{-\alpha}$ where $\alpha=0.155\pm0.035_{stat}{-0.037}_{sys}$) 
{deviates from the synchrotron nebula, where the size shrinks with $E_{Sy}^{-0.085}$. Possible explanations have been considered (transition from Thomson to Klein-Nishina regime and a non-uniform magnetic field). While the former explanation appears to be un-realistic, the latter is a well-known feature of the downstream flow as expected for the Crab nebula \citep{kennel_magnetohydrodynamic_1984}.}

\acknowledgments
PKHY acknowledges the support of the DFG under the research grant 	HO 3305/4-1. We thank J. Hahn and M. Holler for useful discussions. {We thank the anonymous referee for very useful comments which helped to improve the manuscript.}

%

\vspace{5mm}
\facilities{}


\software{fermipy \citep{Wood_Fermipy_2017} and Fermi Science Tools (v11r5p3)}


\begin{thebibliography}{}
	\expandafter\ifx\csname natexlab\endcsname\relax\def\natexlab#1{#1}\fi
	\providecommand{\url}[1]{\href{#1}{#1}}
	\providecommand{\dodoi}[1]{doi:~\href{http://doi.org/#1}{\nolinkurl{#1}}}
	\providecommand{\doeprint}[1]{\href{http://ascl.net/#1}{\nolinkurl{http://ascl.net/#1}}}
	\providecommand{\doarXiv}[1]{\href{https://arxiv.org/abs/#1}{\nolinkurl{https://arxiv.org/abs/#1}}}
	
	\bibitem[{Abdo {et~al.}(2013)Abdo, Ajello, Allafort, Baldini, Ballet,
		Barbiellini, Baring, Bastieri, Belfiore, Bellazzini, Bhattacharyya, Bissaldi,
		Bloom, Bonamente, Bottacini, Brandt, Bregeon, Brigida, Bruel, Buehler,
		Burgay, Burnett, Busetto, Buson, Caliandro, Cameron, Camilo, Caraveo,
		Casandjian, Cecchi, Çelik, Charles, Chaty, Chaves, Chekhtman, Chen, Chiang,
		Chiaro, Ciprini, Claus, Cognard, Cohen-Tanugi, Cominsky, Conrad, Cutini,
		D'Ammando, de~Angelis, DeCesar, De~Luca, den Hartog, de~Palma, Dermer,
		Desvignes, Digel, Di~Venere, Drell, Drlica-Wagner, Dubois, Dumora, Espinoza,
		Falletti, Favuzzi, Ferrara, Focke, Franckowiak, Freire, Funk, Fusco, Gargano,
		Gasparrini, Germani, Giglietto, Giommi, Giordano, Giroletti, Glanzman,
		Godfrey, Gotthelf, Grenier, Grondin, Grove, Guillemot, Guiriec, Hadasch,
		Hanabata, Harding, Hayashida, Hays, Hessels, Hewitt, Hill, Horan, Hou,
		Hughes, Jackson, Janssen, Jogler, Jóhannesson, Johnson, Johnson, Johnson,
		Johnson, Johnston, Kamae, Kataoka, Keith, Kerr, Knödlseder, Kramer, Kuss,
		Lande, Larsson, Latronico, Lemoine-Goumard, Longo, Loparco, Lovellette,
		Lubrano, Lyne, Manchester, Marelli, Massaro, Mayer, Mazziotta, McEnery,
		McLaughlin, Mehault, Michelson, Mignani, Mitthumsiri, Mizuno, Moiseev,
		Monzani, Morselli, Moskalenko, Murgia, Nakamori, Nemmen, Nuss, Ohno, Ohsugi,
		Orienti, Orlando, Ormes, Paneque, Panetta, Parent, Perkins, Pesce-Rollins,
		Pierbattista, Piron, Pivato, Pletsch, Porter, Possenti, Rainò, Rando,
		Ransom, Ray, Razzano, Rea, Reimer, Reimer, Renault, Reposeur, Ritz, Romani,
		Roth, Rousseau, Roy, Ruan, Sartori, Saz~Parkinson, Scargle, Schulz, Sgrò,
		Shannon, Siskind, Smith, Spandre, Spinelli, Stappers, Strong, Suson,
		Takahashi, Thayer, Thayer, Theureau, Thompson, Thorsett, Tibaldo, Tibolla,
		Tinivella, Torres, Tosti, Troja, Uchiyama, Usher, Vandenbroucke, Vasileiou,
		Venter, Vianello, Vitale, Wang, Weltevrede, Winer, Wolff, Wood, Wood, Wood,
		\& Yang}]{abdo_second_2013}
	Abdo, A.~A., Ajello, M., Allafort, A., {et~al.} 2013, The Astrophysical Journal
	Supplement Series, 208, 17, \dodoi{10.1088/0067-0049/208/2/17}
	
	\bibitem[{{Aharonian et al.}(2004)}]{aharonian_et_al._crab_2004}
	{Aharonian et al.} 2004, The Astrophysical Journal, 614, 897,
	\dodoi{10.1086/423931}
	
	\bibitem[{Ansoldi {et~al.}(2016)Ansoldi, Antonelli, Antoranz, Babic, Bangale,
		Barres~de Almeida, Barrio, Becerra~González, Bednarek, Bernardini, Biasuzzi,
		Biland, Blanch, Bonnefoy, Bonnoli, Borracci, Bretz, Carmona, Carosi, Colin,
		Colombo, Contreras, Cortina, Covino, Da~Vela, Dazzi, De~Angelis, De~Caneva,
		De~Lotto, de~Oña~Wilhelmi, Delgado~Mendez, Di~Pierro, Dominis~Prester,
		Dorner, Doro, Einecke, Eisenacher~Glawion, Elsaesser, Fernández-Barral,
		Fidalgo, Fonseca, Font, Frantzen, Fruck, Galindo, García~López,
		Garczarczyk, Garrido~Terrats, Gaug, Godinović, González~Muñoz, Gozzini,
		Hanabata, Hayashida, Herrera, Hirotani, Hose, Hrupec, Hughes, Idec,
		Kellermann, Knoetig, Kodani, Konno, Krause, Kubo, Kushida, La~Barbera, Lelas,
		Lewandowska, Lindfors, Lombardi, Longo, López, López-Coto, López-Oramas,
		Lorenz, Makariev, Mallot, Maneva, Mannheim, Maraschi, Marcote, Mariotti,
		Martínez, Mazin, Menzel, Miranda, Mirzoyan, Moralejo, Munar-Adrover,
		Nakajima, Neustroev, Niedzwiecki, Nevas~Rosillo, Nilsson, Nishijima, Noda,
		Orito, Overkemping, Paiano, Palatiello, Paneque, Paoletti, Paredes,
		Paredes-Fortuny, Persic, Poutanen, Prada~Moroni, Prandini, Puljak, Reinthal,
		Rhode, Ribó, Rico, Rodriguez~Garcia, Saito, Saito, Satalecka, Scalzotto,
		Scapin, Schultz, Schweizer, Shore, Sillanpää, Sitarek, Snidaric,
		Sobczynska, Stamerra, Steinbring, Strzys, Takalo, Takami, Tavecchio,
		Temnikov, Terzić, Tescaro, Teshima, Thaele, Torres, Toyama, Treves, Ward,
		Will, \& Zanin}]{ansoldi_teraelectronvolt_2016}
	Ansoldi, S., Antonelli, L.~A., Antoranz, P., {et~al.} 2016, Astronomy and
	Astrophysics, 585, A133, \dodoi{10.1051/0004-6361/201526853}
	
	\bibitem[{Atoyan \& Aharonian(1996)}]{atoyan_mechanisms_1996}
	Atoyan, A.~M., \& Aharonian, F.~A. 1996, Monthly Notices of the Royal
	Astronomical Society, 278, 525, \dodoi{10.1093/mnras/278.2.525}
	
	\bibitem[{Bietenholz {et~al.}(2004)Bietenholz, Hester, Frail, \&
		Bartel}]{bietenholz_crab_2004}
	Bietenholz, M.~F., Hester, J.~J., Frail, D.~A., \& Bartel, N. 2004, The
	Astrophysical Journal, 615, 794, \dodoi{10.1086/424653}
	
	\bibitem[{Buehler {et~al.}(2012)Buehler, Scargle, Blandford, Baldini, Baring,
		Belfiore, Charles, Chiang, D'Ammando, Dermer, Funk, Grove, Harding, Hays,
		Kerr, Massaro, Mazziotta, Romani, Saz~Parkinson, Tennant, \&
		Weisskopf}]{buehler_gamma-ray_2012}
	Buehler, R., Scargle, J.~D., Blandford, R.~D., {et~al.} 2012, The Astrophysical
	Journal, 749, 26, \dodoi{10.1088/0004-637X/749/1/26}
	
	\bibitem[{Bühler \& Blandford(2014)}]{buhler_surprising_2014}
	Bühler, R., \& Blandford, R. 2014, Reports on Progress in Physics, 77, 066901,
	\dodoi{10.1088/0034-4885/77/6/066901}
	
	\bibitem[{de~Jager \& Harding(1992)}]{de_jager_expected_1992}
	de~Jager, O.~C., \& Harding, A.~K. 1992, The Astrophysical Journal, 396, 161,
	\dodoi{10.1086/171706}
	
	\bibitem[{Dubner {et~al.}(2017)Dubner, Castelletti, Kargaltsev, Pavlov,
		Bietenholz, \& Talavera}]{dubner_morphological_2017}
	Dubner, G., Castelletti, G., Kargaltsev, O., {et~al.} 2017, The Astrophysical
	Journal, 840, 82, \dodoi{10.3847/1538-4357/aa6983}
	
	\bibitem[{{Fermi-LAT Collaboration} \&
		Biteau(2018)}]{fermi-lat_collaboration_search_2018}
	{Fermi-LAT Collaboration}, \& Biteau, J. 2018, ArXiv e-prints, 1804,
	arXiv:1804.08035
	
	\bibitem[{Fraschetti \& Pohl(2017)}]{fraschetti_particle_2017}
	Fraschetti, F., \& Pohl, M. 2017, Monthly Notices of the Royal Astronomical
	Society, 471, 4856, \dodoi{10.1093/mnras/stx1833}
	
	\bibitem[{Gunn \& Ostriker(1971)}]{gunn_motion_1971}
	Gunn, J.~E., \& Ostriker, J.~P. 1971, The Astrophysical Journal, 165, 523,
	\dodoi{10.1086/150919}
	
	\bibitem[{Hester(2008)}]{hester_crab_2008}
	Hester, J.~J. 2008, Annual Review of Astronomy and Astrophysics, 46, 127,
	\dodoi{10.1146/annurev.astro.45.051806.110608}
	
	\bibitem[{Hillas {et~al.}(1998)Hillas, Akerlof, Biller, Buckley, Carter-Lewis,
		Catanese, Cawley, Fegan, Finley, Gaidos, Krennrich, Lamb, Lang, Mohanty,
		Punch, Reynolds, Rodgers, Rose, Rovero, Schubnell, Sembroski, Vacanti,
		Weekes, West, \& Zweerink}]{hillas_spectrum_1998}
	Hillas, A.~M., Akerlof, C.~W., Biller, S.~D., {et~al.} 1998, The Astrophysical
	Journal, 503, 744, \dodoi{10.1086/306005}
	
	\bibitem[{Holler {et~al.}(2017)Holler, Berge, Hahn, Khangulyan, Parsons, \&
		{for the H. E. S. S. collaboration}}]{holler_advanced_2017}
	Holler, M., Berge, D., Hahn, J., {et~al.} 2017, ArXiv e-prints, 1707,
	arXiv:1707.04196
	
	\bibitem[{Kennel \& Coroniti(1984)}]{kennel_magnetohydrodynamic_1984}
	Kennel, C.~F., \& Coroniti, F.~V. 1984, The Astrophysical Journal, 283, 710,
	\dodoi{10.1086/162357}
	
	\bibitem[{Kniffen {et~al.}(1974)Kniffen, Hartman, Thompson, Bignami, \&
		Fichtel}]{kniffen_gamma_1974}
	Kniffen, D.~A., Hartman, R.~C., Thompson, D.~J., Bignami, G.~F., \& Fichtel,
	C.~E. 1974, Nature, 251, 397, \dodoi{10.1038/251397a0}
	
	\bibitem[{Lobanov {et~al.}(2011)Lobanov, Horns, \& Muxlow}]{lobanov_vlbi_2011}
	Lobanov, A.~P., Horns, D., \& Muxlow, T. W.~B. 2011, Astronomy and
	Astrophysics, 533, A10, \dodoi{10.1051/0004-6361/201117082}
	
	\bibitem[{Martín {et~al.}(2012)Martín, Torres, \&
		Rea}]{martin_time-dependent_2012}
	Martín, J., Torres, D.~F., \& Rea, N. 2012, Monthly Notices of the Royal
	Astronomical Society, 427, 415, \dodoi{10.1111/j.1365-2966.2012.22014.x}
	
	\bibitem[{Meyer {et~al.}(2010)Meyer, Horns, \& Zechlin}]{meyer_crab_2010}
	Meyer, M., Horns, D., \& Zechlin, H.-S. 2010, Astronomy and Astrophysics, 523,
	A2, \dodoi{10.1051/0004-6361/201014108}
	
	\bibitem[{Rees(1971)}]{rees_implications_1971}
	Rees, M.~J. 1971, Nature Physical Science, 230, 55,
	\dodoi{10.1038/physci230055a0}
	
	\bibitem[{Spitkovsky \& Arons(2004)}]{spitkovsky_time_2004}
	Spitkovsky, A., \& Arons, J. 2004, The Astrophysical Journal, 603, 669,
	\dodoi{10.1086/381568}
	
	\bibitem[{Trimble(1968)}]{trimble_motions_1968}
	Trimble, V. 1968, The Astronomical Journal, 73, 535, \dodoi{10.1086/110658}
	
	\bibitem[{Volpi {et~al.}(2008)Volpi, Del~Zanna, Amato, \&
		Bucciantini}]{volpi_non-thermal_2008}
	Volpi, D., Del~Zanna, L., Amato, E., \& Bucciantini, N. 2008, Astronomy and
	Astrophysics, 485, 337, \dodoi{10.1051/0004-6361:200809424}
	
	\bibitem[{Weekes {et~al.}(1989)Weekes, Cawley, Fegan, Gibbs, Hillas, Kowk,
		Lamb, Lewis, Macomb, Porter, Reynolds, \& Vacanti}]{weekes_observation_1989}
	Weekes, T.~C., Cawley, M.~F., Fegan, D.~J., {et~al.} 1989, The Astrophysical
	Journal, 342, 379, \dodoi{10.1086/167599}
	
	\bibitem[{{Wood} {et~al.}(2017){Wood}, {Caputo}, {Charles}, {Di Mauro},
		{Magill}, {Perkins}, \& {Fermi-LAT Collaboration}}]{Wood_Fermipy_2017}
	{Wood}, M., {Caputo}, R., {Charles}, E., {et~al.} 2017, International Cosmic
	Ray Conference, 301, 824.
	\newblock \doarXiv{1707.09551}
	
\end{thebibliography}
\end{document}


\title{Erratum: ``The Energy-dependent $\gamma$-ray Morphology of the Crab Nebula Observed with the \emph{Fermi} Large Area Telescope" (2019, ApJ, 875, 123)}

\correspondingauthor{Paul K. H. Yeung}
\email{kin.hang.yeung@desy.de}

\author[0000-0003-3476-022X]{Paul K. H. Yeung}
\affil{Institute for Experimental Physics, Department of Physics, University of Hamburg, Luruper Chaussee 149, D-22761 Hamburg, Germany}

\author[0000-0003-1945-0119]{Dieter Horns}
\affil{Institute for Experimental Physics, Department of Physics, University of Hamburg, Luruper Chaussee 149, D-22761 Hamburg, Germany}

\section*{} \label{sec:intro}

In our original article \citet{Yeung_Morphology_2019},  the superimposed spatial model assigned to the whole Crab (i.e. a superposition of a point component [pulsar] and an extended component [nebula]) was following the analysis scheme of \citet{fermi-lat_collaboration_search_2018}. Nevertheless, there is a deficiency of our $\gamma$-ray extension measurements for the Crab Nebula, which is stemming from the Crab pulsar's model. There is not only deviation in spectral modelling for the pulsar component, but also inaccuracy in convolving the pulsar's spatial morphology (i.e. the point source morphology) with the PSF. According to Figure 3 of our \citet{yeung_pulsar_2020}, the pulsar's differential flux accounts for $\sim$(15--72)\% of the Crab's total differential flux from 5 to 40~GeV. Therefore, the aforementioned systematic effect associated with the pulsar component was particularly large for the 5--40~GeV extension measurements reported in our original article.

With regards to this issue, we perform follow-up checks for the 5--40~GeV extension sizes by selecting only off-pulse phase data for analyses and, accordingly, removing the pulsar component from the source model. We adopt the timing solution of the Crab pulsar provided by M. Kerr. We define the off-pulse phase in the same way as in our \citet{yeung_dimming_2019}. We perform binned maximum-likelihood analyses, with an angular bin size of $0.01^\circ$, on P8R3 data accumulated over the first 10 years of observations. We adopt the 4FGL point source catalogue \citep[][]{Fermi_Fourth_2019}, the Galactic diffuse model ``gll$_-$iem$_-$v07" and the isotropic diffuse model ``iso$_-$P8R2$_-$CLEAN$_-$V6$_-$v07". Other details of data reduction criteria and analysis procedures roughly follow our original article.

It turns out that the disk radius of the Crab Nebula is revised to be $0.041^{+0.016}_{-0.025}$, $0.070^{+0.008}_{-0.009}$ and $0.042 ^{+0.009}_{-0.011}$~deg in 5--10, 10--20 and 20--40~GeV respectively. Moreover, the revised disk radii in the combined 5--20~GeV segment and the broad 5--500~GeV band are $0.064^{+0.007}_{-0.008}$ and  $0.047^{+0.005}_{-0.006}$~deg respectively.

\begin{figure}
	\plotone{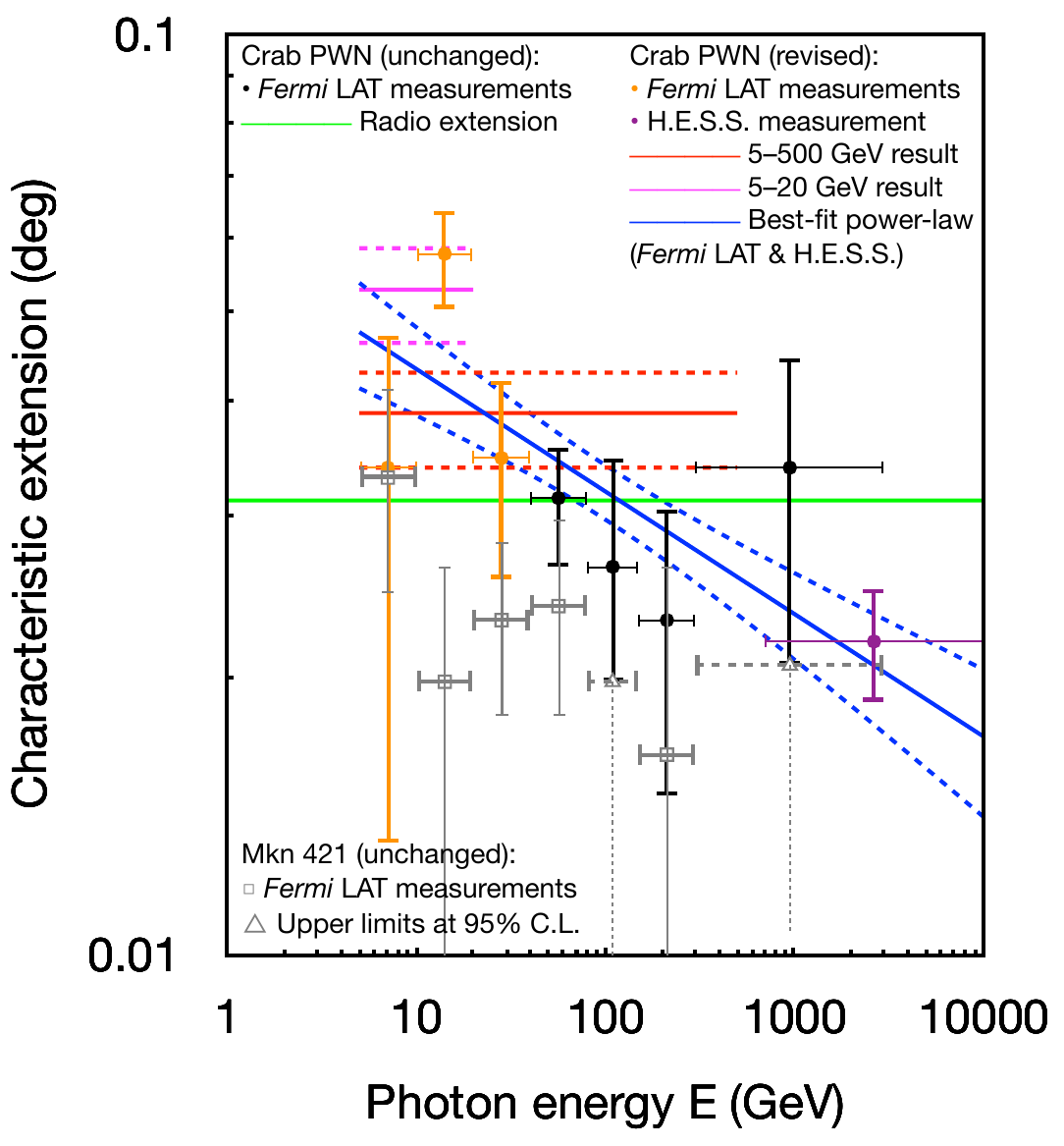}
	\caption{Spectral distribution of characteristic  extensions (defined as the 68\% containment radius $R_{68}$;  either the uniform-disk radius multiplied by $\sqrt{0.68}$ or the Gaussian rms width multiplied by $\sqrt{2 ln(1/0.32)}$) of the Crab PWN (modified from Figure 6 of our original paper). The orange bins are our revised LAT results in this erratum, and the black bins are the LAT results taken from our original article.  The H.E.S.S. bin (purple) is revised to be the result of \citet{HESS_ext_2020}, where the best-fit value remains the same as before but the uncertainty becomes smaller. The revised $R_{68}$ of the energy-independent disk morphologies fit to the 5--500 and  5--20~GeV datasets are plotted in red and  magenta respectively.  The blue line is revised to be the power-law fit to all orange, black and purple bins of this modified figure: $R_{68}=(0.0317\pm0.0022)(\frac{E}{103.8~\mathrm{GeV}})^{-0.133\pm0.037}$ deg, where the prefactor and index are decorrelated.  \label{FermiHESS_rev}} 
\end{figure}

As demonstrated in Figure~\ref{FermiHESS_rev}, these revised results do not  throw doubt on the conclusions on excess 5--20~GeV extensions (relative to the radio extension) and on energy-dependent shrinking in our original paper.

